\documentclass[useAMS]{mn2e}
\usepackage{graphicx}
\usepackage{epsfig}
\usepackage{amssymb}
\usepackage{lscape}
\usepackage{ulem}
\usepackage{txfonts}

\def\gal{SDSS J140737.17+442856.2}

\def\chisq{$\chi^2$}
\def\chisqdof{\chi^2\textrm{/dof}}
\def\nh{N_{H}}
\def\cmsq{cm$^{-2}$}

\def\ergcms{erg~cm$^{-2}$~s$^{-1}$}
\def\ergs{erg~s$^{-1}$}

\def\Fxtwoten{F_{\textrm{2-10~keV}}}
\def\Lxtwoten{L_{\textrm{2-10~keV}}}

\def\chandra{\textit{Chandra}}

\def\kms{km~s$^{-1}$}
\def\cm{cm$^{-2}$}

\def\c2s{C\,{\sc ii}$^{\star}$}

\title[A new dual AGN] {Discovery of a dual active galactic nucleus with $\sim$ 8 kpc separation.}

\author[Ellison et al.] {Sara L. Ellison$^1$, Nathan J. Secrest$^{2,3}$, J. Trevor Mendel$^4$, Shobita Satyapal$^5$, Luc Simard$^6$\\
$^1$ Department of Physics \& Astronomy, University of Victoria, Finnerty Road, Victoria, British Columbia, V8P 1A1, Canada.\\
$^2$ National Academy of Sciences NRC Research Associate, 2101 Constitution Ave. NW, Washington, DC 20418, USA\\
$^{3}$ Resident at Naval Research Laboratory, 4555 Overlook Ave. SW, Washington, DC 20375, USA.\\
$^4$ Max-Planck-Institut fur Extraterrestrische Physik, Giessenbachstrasse, D-85748 Garching, Germany.\\
$^5$ George Mason University, Department of Physics \& Astronomy, MS 3F3, 4400 University Drive, Fairfax, VA 22030, USA\\
$^6$ National Research Council of Canada, Herzberg Institute of Astrophysics, 5071 West Saanich Road, Victoria, British Columbia, V9E 2E7, Canada
}

\begin{document}

\maketitle

\begin{abstract}
Targeted searches for dual active galactic nuclei (AGN), with separations 1 -- 10 kpc, have yielded relatively few successes.  A recent pilot survey by Satyapal et al. has demonstrated that mid-infrared (mid-IR) pre-selection has the potential to significantly improve the success rate for dual AGN confirmation in late stage galaxy mergers.  In this paper, we combine mid-IR selection with spatially resolved optical AGN diagnostics from the Mapping Nearby Galaxies at Apache Point Observatory (MaNGA) survey to identify a candidate dual AGN in the late stage major galaxy merger \gal\ at $z=0.143$.  The nature of the dual AGN is confirmed with $Chandra$ X-ray observations that identify two hard X-ray point sources with intrinsic (corrected for absorption) 2-10 keV luminosities of 4$\times10^{41}$ and 3.5$\times10^{43}$ erg/s separated by 8.3 kpc.   The neutral hydrogen absorption ($\sim$ 10$^{22}$ \cm) towards the two AGN is lower than in duals selected solely on their mid-IR colours, indicating that strategies that combine optical and mid-IR diagnostics may complement techniques that identify the highly obscured dual phase, such as at high X-ray energies or mid-IR only.
\end{abstract}


\section{Introduction}

Galaxy mergers represent a cornerstone of the hierarchical picture of structure formation.  In the nearby universe, a few percent of galaxies are engaged in a major interaction and these can result in dramatic transformations in morphology (e.g. Patton et al. 2016), star formation rate (e.g. Ellison et al. 2008) and chemical composition (e.g. Scudder et al. 2012).  In addition to the build-up of stellar mass, galaxy mergers are presumed to also result in the coalescence of the central supermassive black holes, leading to a coordinated growth of the two components (Jahnke \& Maccio  2011).  Binary black holes should therefore be abundant in late stage galaxy mergers. These merging supermassive black holes are predicted to produce strong gravitational wave signals, and constraining their frequency is therefore important for future experiments such as the Laser Interferometer Space Antenna (LISA) project (e.g.  Sesana et al. 2004; Khan et al., 2016).  Fortunately, the galaxy merger process triggers inward gas flows that lead to AGN (Ellison et al. 2011, 2013; Khabiboulline et al. 2014; Lackner et al. 2014), potentially making accreting black hole pairs ripe for detection. 

In reality, the hunt for binary (separation $<$ 1 kpc) or dual (separation 1 -- 10 kpc) AGN has been a frustrating one.  The identification of AGN pairs began with serendipitous discoveries (e.g. Komossa et al. 2003; Ballo et al. 2004; Bianchi et al. 2008), providing encouragement that such objects existed. Several teams then set about making more systematic searches for dual AGN, with the hope of not just significantly increasing the known numbers but also performing a systematic characterization of the black hole merger process. These searches often leveraged pre-selection from large galaxy surveys, targeting, for example, galaxies with double-peaked emission lines (e.g. Comerford et al. 2012), or visually classified mergers (Teng et al. 2012).  Although these searches have yielded some new dual AGN identifications (e.g. Comerford et al. 2015), the confirmation rate is low, meaning that large amounts of telescope time are required for a relatively low yield (Fu et al., 2012).  A new approach appears to be necessary if dual AGN are to be identified more efficiently and in larger numbers.

\medskip

Evidence is now building that the AGN in galaxy mergers are more obscured than those in isolated galaxies (Liu et al. 2013;  Kocevski et al. 2015; Ricci et al. 2017; Satyapal et al. 2017).  In particular, AGN identified via their mid-infrared (mid-IR) colours are prevalent in mergers  (Satyapal et al. 2014; Ellison, Patton \& Hickox 2015).  Simulations of interacting galaxies indicate that such AGN occur preferentially at the late stages in a merger (close to final coalescence) when the AGN is both bolometrically dominant \textit{and} there is a high obscuring column density of gas and dust (Blecha et al., in prep).   These combined works have demonstrated the excellent potential of using mid-IR colours from the Wide Field Infrared Survey Explorer (WISE) satellite to identfy AGN in late stage mergers.  On this basis, we have recently completed a pilot $Chandra$ search for dual AGN in 6 late stage galaxy mergers with red WISE colours ($W1 - W2>0.5$).  In Satyapal et al. (2017) we have presented 4 new duals identified out of the 6 systems observed, demonstrating that WISE colour selection is a highly effective way of identifying dual AGN candidates.  Notably, the 4 new dual AGN all have low X-ray (2-10 keV) to mid-IR (12 $\mu$m) fluxes relative to the local $Swift/BAT$ AGN sample (Ricci et al., 2015) indicating significant hydrogen column densities along the line of sight.  

Here, we extend the strategy of WISE colour selection for the identification of dual AGN, by searching for candidates that have spatially resolved spectroscopy in the public domain.  Specifically, we report the detection of a new dual AGN with physical separation of 8.3 kpc at $z=0.143$ identified by combining WISE colour selection with integral field spectroscopy from the Mapping Nearby Galaxies at Apache Point Observatory (MaNGA) survey, and final confirmation with the $Chandra$ X-ray Observatory.

\section{Target selection and data analysis}

A targeted search for rare astrophysical objects benefits from judicial sifting through large public datasets.  In this work, we take advantage of two major public astronomical surveys: the WISE All Sky Survey (Wright et al. 2010) and the Sloan Digital Sky Survey (SDSS), with a particular emphasis on the MaNGA survey (Bundy et al. 2015).

\vspace{-0.5cm}

\subsection{Target selection: WISE + SDSS Legacy data}

The MaNGA survey is one of three projects within SDSS-IV.  By bundling together the individual 2$\arcsec$ fibres of the twin Baryon Oscillation Spectroscopic Survey (BOSS) spectrographs into hexagonal integral field units (IFUs), and employing a dithering strategy to fill in the gaps between fibres, a continuous spectral map can be obtained (Law et al. 2015).  Galaxies are selected from the SDSS Main Galaxy (Legacy) sample and observed with a size-matched IFU that varies in diameter from 12$\arcsec$ (19 fibres) to 32$\arcsec$ (127 fibres). In July 2016 the SDSS Data Release (DR) 13 included the first MaNGA data release of $\sim$ 1400 galaxies.

We cross-matched the galaxies in the MaNGA DR13 with the WISE All Sky
Survey 
requiring a positional match better than 6$\arcsec$.  We use a $W1 - W2 > 0.8$ colour cut (Stern et al. 2012) to identify potential AGN.  Blecha et al. (in prep) have shown that such red WISE colours occur preferentially in the late stage of a merger when the nuclei are within a few kpc of one another.  We identify a single galaxy in the first MaNGA data release that passes the $W1-W2>0.8$ criterion: SDSS J140737.17+442856.2 (objID 588298663581974759).  With a  $W1-W2=0.84$ and  $W2-W3=2.44$, this source is also located within the 3-colour AGN selection box of Jarrett et al. (2011).

\begin{figure}
\centering
\includegraphics[width=6.5cm,angle=0]{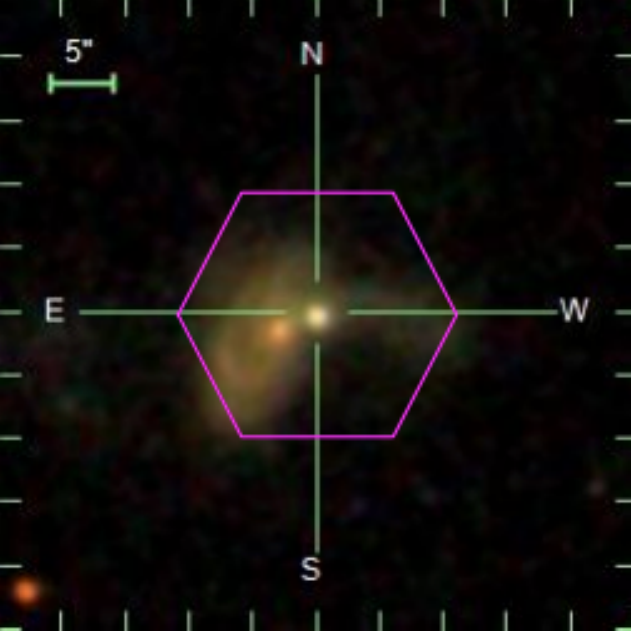}
\includegraphics[width=7cm,angle=0]{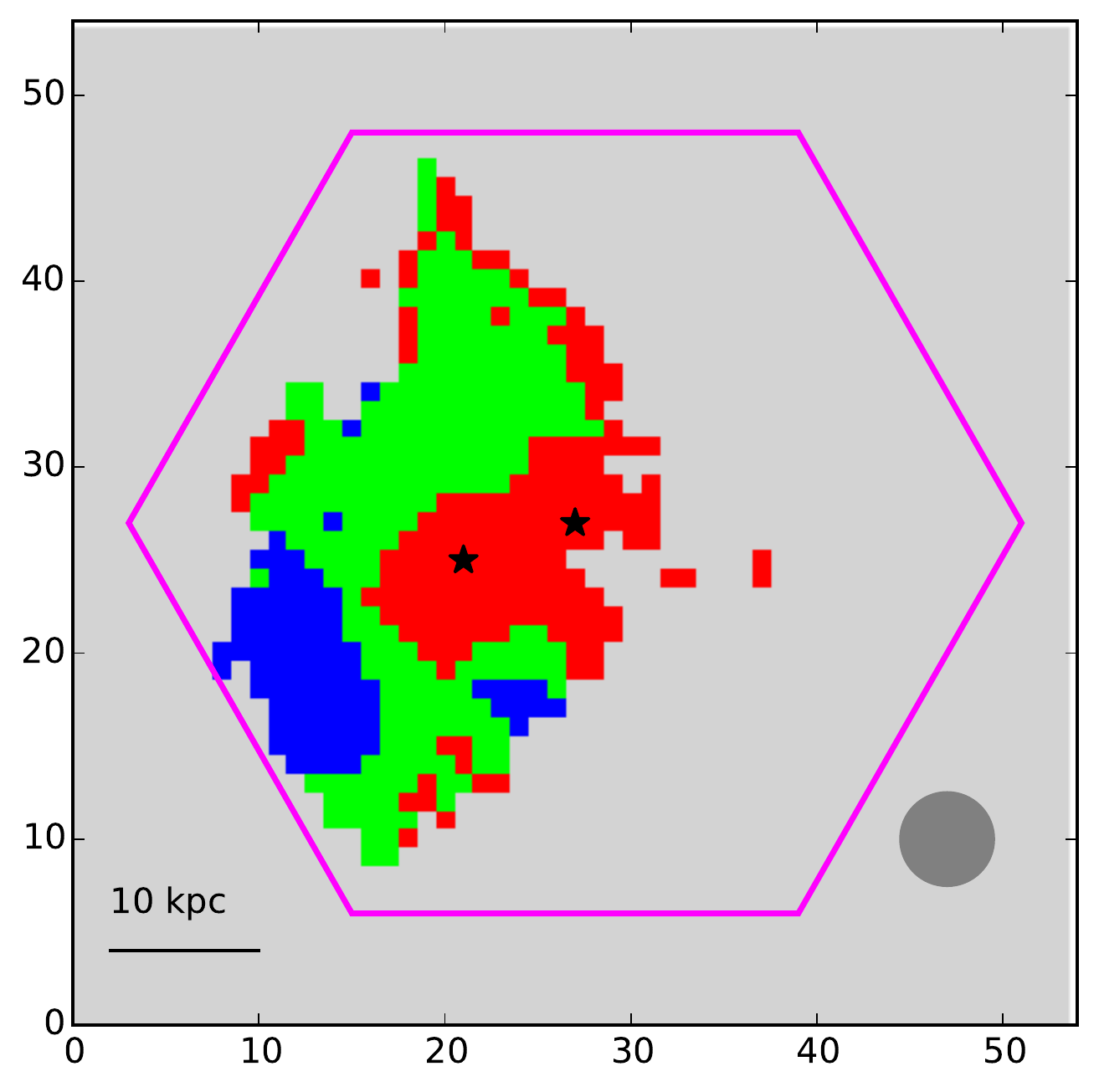}
\caption{Top panel: SDSS $gri$ image of galaxy SDSS J140737.17+442856.2, centred on the western nucleus, which corresponds to the position of the DR7 Legacy spectrum.  The position of the WISE source is centred on the eastern nucleus, which is separated by $\sim$3$\arcsec$.  Both nuclei have separate single fibre spectra from the SDSS that classify them as AGN.  Lower panel: Line ratio diagnostic map for SDSS J140737.17+442856.2 from single component fit to the MaNGA data.  Spaxels with S/N$>$3 in [NII] and H$\alpha$ are colour-coded by their classification on the [NII]/H$\alpha$ vs [OIII]/H$\beta$ emission line ratio diagram: red, blue and green indicate gas dominated by AGN, star formation or a combination (respectively). The MaNGA IFU footprint is shown with the magenta hexagon, the $2\arcsec\hspace{-1.5mm}.\hspace{0.75mm}5$ FWHM PSF with the dark grey circle and the positions of SDSS Legacy and BOSS fibre spectra (centred on the galactic nuclei) are shown with black stars. }
\label{image}
\end{figure}

Based on visual inspection of the SDSS image (Fig. \ref{image}, upper panel), SDSS J140737.17+442856.2 appears to be a late stage galaxy merger in which two stellar nuclear are separated by $\sim3\arcsec$.   A registered comparison of source positions shows that the SDSS DR7 spectrum is associated with the western nucleus ($z_{\rm DR7}$ = 0.14289), whereas the WISE position is centred on the eastern nucleus.   We use (narrow) emission lines from the MPA/JHU spectral fits (Brinchmann et al. 2004) to confirm that the western source is a Type II AGN on the [NII]/H$\alpha$ vs [OIII]/H$\beta$ emission line ratio diagram (Kewley et al. 2001).  The eastern target also has an SDSS spectrum - it was observed as part of BOSS (part of SDSS-III), in which it is classified as a QSO ($z_{\rm BOSS}$ = 0.14301).  The velocity difference between the two stellar nuclei is therefore only $\sim$ 30 \kms.

\subsection{Stellar masses}

Mendel et al. (2014) have derived stellar masses for $\sim$ 660,000 SDSS galaxies based on the bulge+disk photometry presented in Simard et al. (2011).  Mendel et al. (2014) determine a stellar mass for the western source of log (M$_{\star}$/M$_{\odot}$) = 11.08.  The eastern source was not included in the Mendel et al. (2014) catalog, which was based on spectroscopic targets in the DR7.  However, the eastern source does have photometry available in Simard et al. (2011), which includes photometric objects with 14 $< r <$ 18.  Using identical methods as Mendel et al. (2014) yields a total stellar mass for the eastern source of log (M$_{\star}$/M$_{\odot}$) = 11.15, indicating that the merger is between approximately equal mass galaxies.  However, because a bulge+disk model is unlikely to accurately represent the internal structure in this late stage merger, we re-computed the $ugriz$ photometry by simply summing up the flux associated with each galaxy, manually drawing a division between the two nuclei.  Masses were computed using the same models as in Mendel et al. (2014).  The stellar masses derived from this aperture photometry approach were log (M$_{\star}$/M$_{\odot}$) = 10.84 and 11.02 for the western and eastern galaxies, respectively.  Whilst there are drawbacks to both approaches (as shown in Simard et al. 2011, aperture photometry of close pairs can incorrectly allocate flux between the companions), we conclude that \gal\ is likely to be a major merger between approximately equal mass galaxies.

\subsection{MaNGA data}

We model spectra in the MaNGA data cube using a combination of continuum (stellar) templates and Gaussian emission lines.  Our continuum templates are derived from the stellar library of Sanchez-Blazquez et al. (2009) using the diffusion K-means algorithm described by Richards et al. (2009); we found that a total of 35 `optimal' templates were sufficient to describe the spectra in the MaNGA datacube. In modeling the emission lines we consider both single- and two-component models.

The strong degeneracies inherent to modeling multiple emission-line components motivates us to consider a Bayesian approach in our analysis.  We use the MultiNest algorithm (Feroz \& Hobson 2007; Feroz et al. 2009, 2013) accessed via the PyMultiNest wrapper (Buchner et al 2014) to compute the evidence, $Z$, for both single- and two-component emission-line models, as well as generate samples from the posterior distribution.  We identify spaxels hosting multiple emission line components as those with a Bayes factor ($\equiv Z_\mathrm{double}/Z_\mathrm{single}$) greater than 5.  We find that, in general, a single emission line component is sufficient to describe the MaNGA data, although there is strong evidence for an additional broader component associated with the eastern nucleus, consistent with the interpretation of the BOSS spectrum\footnote{The broad line component has a relatively low $\sigma \sim$ 200 \kms, indicating that it is tracing a hotter kinematic component of the gas, but not the broad line region around the black hole itself.} .

In the lower panel of Fig. \ref{image} we show the single component emission line diagnostic map of SDSS J140737.17+442856.2 derived from the MaNGA data cube.  After dithering and image reconstruction, the data cube is re-sampled with $0\arcsec\hspace{-1.5mm}.\hspace{0.75mm}5$ spaxels, but the median full width at half maximum (FHWM) of the MaNGA point spread function (PSF) is $\sim$ $2\arcsec\hspace{-1.5mm}.\hspace{0.75mm}5$ (Law et al. 2016), shown by the dark gray circle in the lower right of Fig. \ref{image}. Spaxels are colour-coded according to classification on the [NII]/H$\alpha$ vs [OIII]/H$\beta$ emission line ratio diagram: red indicates an AGN (above the Kewley et al. 2001 line), green indicates a composite (between the Kauffmann et al. 2003 and Kewley et al. 2001 lines) and blue indicates gas dominated by star formation (below the Kauffmann et al. 2003 line).  The positions of the DR7 and BOSS spectra (centred on the galactic nuclei) shown with black stars.  The line ratio map suggests that the gas around both stellar nuclei is photo-ionized by an AGN, a conclusion supported by other diagnostic line ratio maps ([OI]/H$\alpha$ and [SII]/H$\alpha$ vs [OIII]/H$\beta$).  In the two-component fit (not shown) the AGN classification of the eastern source is dominated by the broader component. However, although the MaNGA data (line ratios and morphology) are suggestive of two AGN, each photo-ionizing a region around their respective nuclei, it remains a possibility that a single AGN is responsible for ionizing the whole extended region.  The optimal way to confirm the presence of an AGN is from the detection of a hard X-ray point source.

\vspace{-0.5cm}

\subsection{Chandra Observations and Data Analysis}
\label{subsec:Chandra_Analysis} 

\begin{figure}
	\includegraphics[width=\columnwidth]{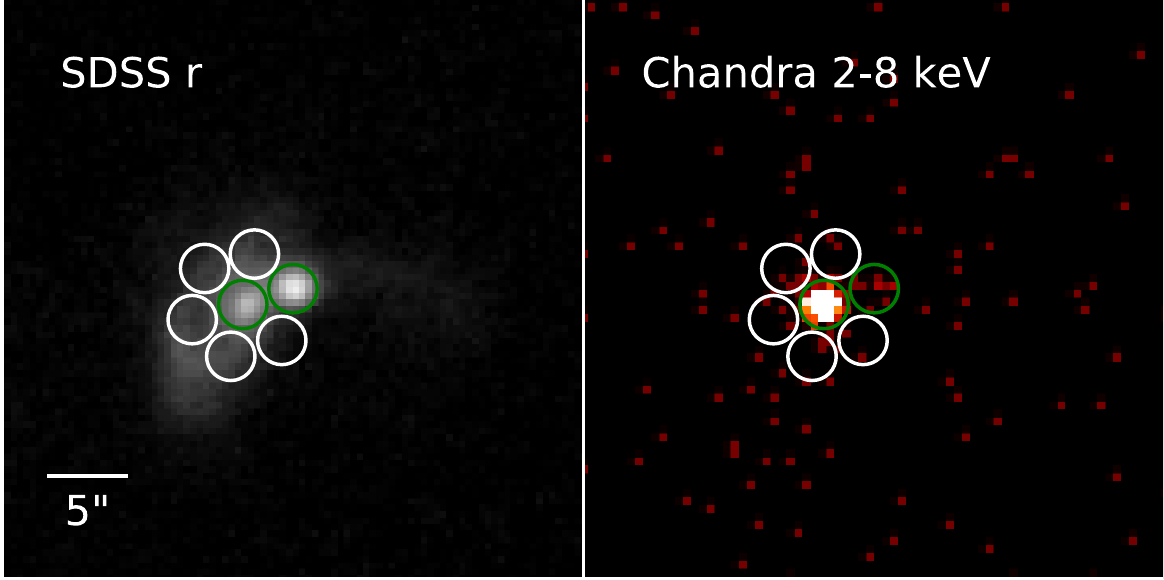}
    \caption{The SDSS $r$-band and \chandra\ 2-8~keV images of \gal, showing the extraction apertures for X1 (eastern source, left green circle) and X2 (western source, right green circle) and the apertures (white) used to determine the significance of X2.}
    \label{fig:rX}
\end{figure}

We obtained a single, 30~ks \chandra\ ACIS-S observation of \gal\ as a Director's Discretionary Time (DDT) program (ObsID=19990; PI: Secrest) on 2017 February 28.  We reprocessed the event and calibration files for this data set using \textsc{ciao} (Fruscione et al. 2006), version 4.9, and CALDB 4.7.3.  We produced filtered event files and images using \texttt{dmcopy}, and extracted counts using \texttt{dmextract}.  We checked for background flares by masking sources in the field with \texttt{wavdetect}; there were no significant ($>3\sigma$) flaring intervals during the observation.  Two X-ray sources, X1 and X2 (separated by $3\arcsec\hspace{-1.5mm}.\hspace{0.75mm}3$), are present in the 2-8~keV image of \gal\ (Figure~\ref{fig:rX}) that correspond to the positions of the eastern and western stellar nuclei in the SDSS image.  By registering sources in the 0.3-8~keV image as determined using \texttt{wavdetect} with their optical counterparts in the SDSS, we determined that the 90\% upper limit on any astrometric offset between the SDSS and \chandra\ data is $0\arcsec\hspace{-1.5mm}.\hspace{0.75mm}2$, confirming that X1 and X2 are indeed the X-ray counterparts to the stellar nuclei. At a redshift of $z=0.143$, and assuming a standard cosmology of $\Omega_{\Lambda} = 0.7$, $\Omega_M = 0.3$, $H_0 = 70$ km/s/Mpc, the source separation of $3\arcsec\hspace{-1.5mm}.\hspace{0.75mm}3$ corresponds to a physical separation of 8.3 kpc.

\begin{table}
	\centering
	\caption{Positions and background-subtracted source counts for X1 and X2.}
	\label{tab:X1X2}
	\begin{tabular}{lccrrr}
		\hline
		X & R.A. & Decl. & 0.3-8~keV  & 2-8~keV  & 0.3-2~keV \\
                  & deg.  & deg.  & (total)    &  (Hard)  & (Soft) \\
		\hline
		1 & 211.90608 & 44.48200 & 1025 & 737 & 288\\
		2 & 211.90487 & 44.48227 &     12 &   10 &      2\\
		\hline
	\end{tabular}
\end{table}

\subsubsection{X1}
\label{subsubsec:X1}

For the brighter source X1, we used the \texttt{wavdetect} source coordinates and extracted counts within a $3\arcsec$ aperture. Using a background annular region centred at X1 with $r_\mathrm{inner}=6\arcsec$ and $r_\mathrm{outer}=20\arcsec$, we find 1025 net (background-subtracted) counts between 0.3 and 8 keV (Table~\ref{tab:X1X2}).  Using \texttt{specextract}, we produced a spectrum of X1, which we grouped using \textsc{grppha}, requiring a minimum grouping of 20 counts for the \chisq\ statistic.  We fit the spectrum using \textsc{xspec}, version 12.9.0 (Arnaud et al. 1996), between 0.3 and 8 keV, modeling the Galactic hydrogen column density as a fixed photoelectric absorption component (\texttt{phabs}) with a value of $8.60\times10^{19}$ \cmsq, calculated using the \textit{Swift} \texttt{nhtot} tool\footnote{http://www.swift.ac.uk/analysis/nhtot/}, which uses the prescription of Willingale et al. (2013).

\begin{figure}
	\includegraphics[width=\columnwidth]{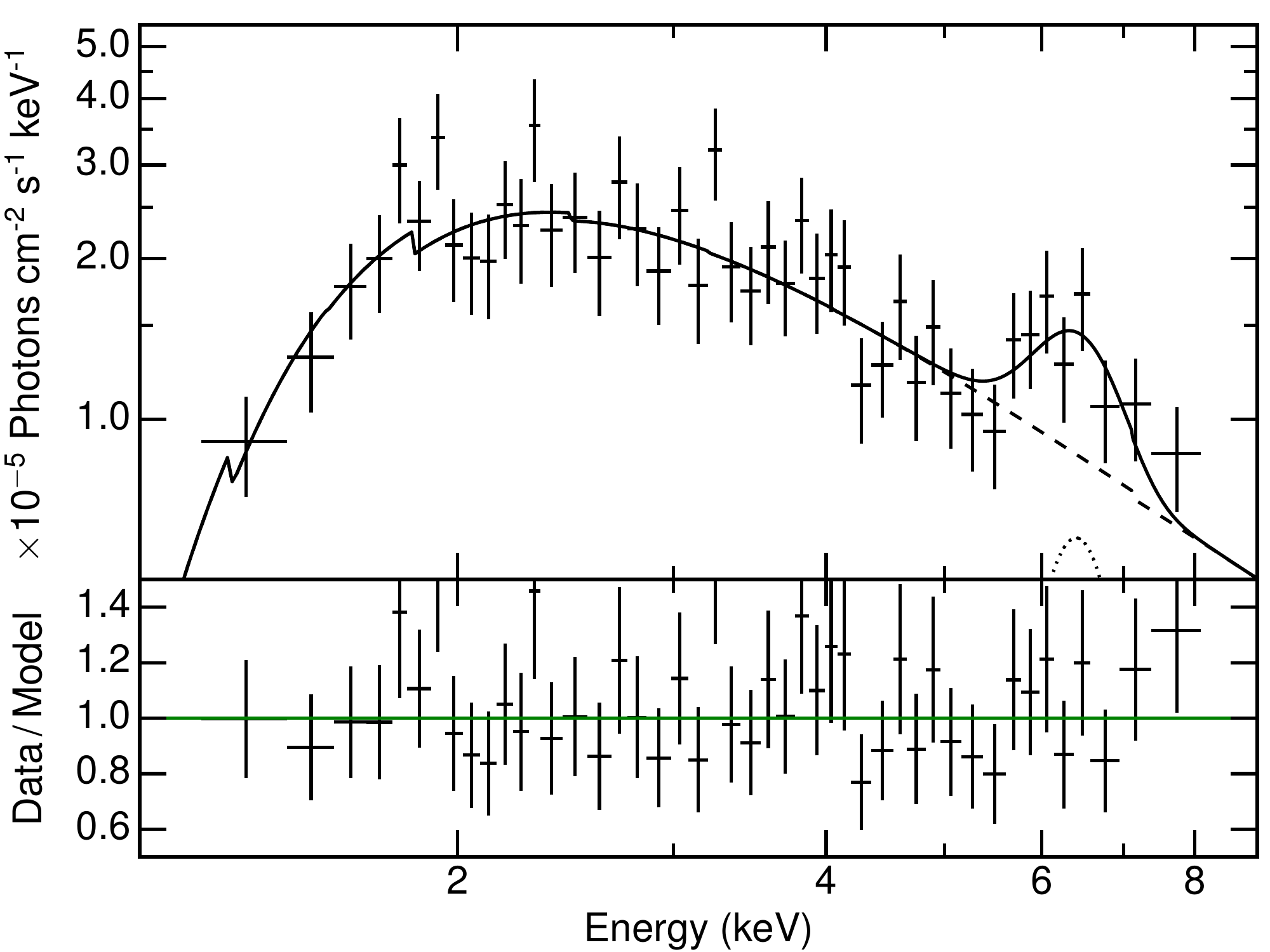}
    \caption{The unfolded rest-frame X-ray spectrum of X1.  The absorbed power-law component is denoted by the dashed line, while the Fe K$\alpha$ line is shown as the dotted line.}
    \label{fig:spectrum}
\end{figure}

We found that the spectrum is well-fit ($\chisqdof=37.26/43$) with an absorbed, redshifted power-law model (\texttt{phabs*zphabs*zpow}) with $\Gamma=1.3^{+0.3}_{-0.2}$ and intrinsic $\nh=(1.6\pm0.4)\times10^{22}$~\cmsq\, where the uncertainties are 90\%.  However, this power-law index $\Gamma$ is flatter than what is typically seen in AGNs (e.g. Marchesi et al., 2016 and references therein), and is apparently due to an excess of X-ray counts between $\sim5$ and 6 keV (observed frame).  Indeed, if we append an Fe~K$\alpha$ line to the model in the form of a Gaussian component, the best-fit power-law index is $\Gamma=1.6^{+0.5}_{-0.4}$, with $\nh=2.0^{+0.6}_{-0.5}\times10^{22}$~\cmsq  (Figure~\ref{fig:spectrum}).  The equivalent width of the Fe~K$\alpha$ line is 0.76 keV.  Using this model, we calculated unabsorbed fluxes using \texttt{cflux}, holding the power-law component normalization fixed to the best-fit value.  The power-law component of X1 has an unabsorbed flux of $\Fxtwoten=6.3^{+0.7}_{-0.8}\times10^{-13}$~\ergcms, corresponding to an \textit{intrinsic} luminosity of $\Lxtwoten=(3.5\pm0.4)\times10^{43}$~\ergs.   Of the previously studied dual AGN, the majority exhibit low (observed) X-ray fluxes relative to either their [OIII] luminosity (e.g. Liu et al. 2013) or their mid-IR luminosity (e.g. Satyapal et al. 2017), consistent with high obscuring columns of $N_H$.  However, with an \textit{observed} $\Lxtwoten= 3.3 \times10^{43}$~\ergs, L$_{12\mu m} = 1.2 \times 10^{44}$ \ergs, and [OIII] luminosity in the central $\sim2\arcsec\hspace{-1.5mm}.\hspace{0.75mm}5$ of the MaNGA datacube of $\sim 5\times 10^{41}$ \ergs, X1 is  consistent with unobscured AGN in $\Lxtwoten$-L$_{12\mu m}$ (Satyapal et al. 2017) and $\Lxtwoten$-L$_{[OIII]}$ (Liu et al. 2013) parameter space.  This is consistent with the lower $N_H$ observed towards X1 than many other dual AGN and late stage mergers, which are frequently Compton thick (log $N_H$ $\gtrsim$ 24; Ricci et al. 2017; Satyapal et al. 2017).  Finally, we note that X1 exhibits the highest observed X-ray luminosity of any confirmed dual AGN to date from $Chandra$ data (see e.g. the compilation in Satyapal et al. 2017).

\subsubsection{X2}
\label{subsubsec:X2}

For X2, we centred the 3$\arcsec$ aperture on the NASA/IPAC Extragalactic Database (NED) position for the optical counterpart SDSS J140737.16+442856.1 with the same background annular region as defined for X1, yielding 12 net counts between 0.3 and 8 keV for X2 (Table~\ref{tab:X1X2}). Due to its relative faintness and proximity to X1, we tested for the possibility that X2 is an artifact of X1 by creating five additional apertures around X1 with the same proximity as X2  (white circles in Figure~\ref{fig:rX}).  We find that in none of the additional apertures are there more than 7 net counts between 0.3 and 8 keV, with an average of 3 counts per aperture.  To remain conservative, we subtracted 3 counts from the 12 measured for X2 to correct for contamination of X2 by X1. The average contamination in the soft 0.3-2~keV band from X2 is $\sim0$~counts.
  

We set constraints on the spectrum of X2 using the Bayesian Estimation of Hardness Ratios code (Park et al. 2006), which gives a mean $\mathrm{(H-S)/(H+S)}=0.58^{+0.42}_{-0.36}$, where the uncertainties are 90\%.  For an AGN power-law spectrum with $\Gamma=1.8$, this corresponds to $\nh\sim3.4\times10^{22}$~\cmsq\ (using \textsc{pimms}\footnote{http://cxc.harvard.edu/toolkit/pimms.jsp}), similar to X1.  For this estimate of $\nh$, the unabsorbed flux is $\Fxtwoten\sim7.3\times10^{-15}$~\ergcms\, corresponding to an intrinsic luminosity of $\Lxtwoten\sim4.0\times10^{41}$~\ergs.  The SFR of this galaxy (taken from the MPA/JHU catalog) is 1.4 M$_{\odot}$ yr$^{-1}$, leading to a predicted flux from X-ray binaries (from Table 4 in Lehmer et al. 2010) that is approximately two orders of magnitude below the observed $\Lxtwoten$. We emphasize that, given the upper bound on the hardness ratio, the upper limit on $\nh$ is relatively unbounded, so the intrinsic luminosity of X2 may be much larger.  Conversely it is unlikely that the intrinsic luminosity is significantly lower than this value, as the lower bound on the hardness ratio corresponds to $\nh\sim1.6\times10^{22}$~\cmsq.  The combined evidence of the optical emission line ratios, hardness ratio and X-ray luminosity lead us to conclude that X2 is also an AGN.

\section{Discussion}

We present the discovery of a dual AGN separated by 8.3 kpc in SDSS J140737.17+442856.2, a major merger between two massive galaxies of log (M$_{\star}$/M$_{\odot}$), selected by combining a red WISE colour threshold ($W1-W2>0.8$) and traditional optical emission line diagnostics.  This combination of criteria selects a late stage merger which has one nucleus with $W1-W2=0.84$ (eastern nucleus in Fig \ref{image}) and the other nucleus that is classified as an optical (Type II) AGN (western nucleus in Fig \ref{image}).  The nature of the dual AGN is confirmed by $Chandra$ X-ray observations, the two nuclei exhibiting intrinsic 2-10 keV luminosities of  4$\times10^{41}$ (western nucleus) and 3.5$\times10^{43}$ (eastern nucleus) \ergs; the latter of these has the highest X-ray luminosity measured by $Chandra$ in a confirmed dual AGN to date.

Previous to our pilot campaign that uses WISE pre-selection (Satyapal et al. 2017) there were only nine dual ($1 < r < 10$ kpc) AGN with X-ray confirmation.  The addition of 5 new duals confirmed by $Chandra$ (4 in Satyapal et al. 2017 and one presented here) represents a significant new haul of close AGN pairs -- increasing the previously known sample by 50 percent.  The WISE-only selection of Satyapal et al. (2017) preferentially identifies highly absorbed (possibly Compton thick) AGN, whereas the inferred neutral hydrogen columns for the AGN pair presented here are $\sim$ 10$^{22}$ \cm.  The lower absorption is consistent with the  detection of AGN signatures in the optical.
Conversely, the dual AGN in Satyapal et al. (2017) are classified optically as star forming, consistent with more highly obscured nuclei.  Ricci et al. (2017) have also found a predominance of obscured AGN in late stage mergers:  of the late stage mergers in their local sample, 65 percent are Compton thick and all have $\log N_H > 23$.   Our results show that whilst observations (e.g. Kocevski et al. 2015; Ricci et al. 2017) and theory (Blecha et al. in prep) alike suggest that obscuration is high in late stage mergers, not all AGN in mergers are enshrouded by high columns of gas and dust.  

Resolved spectroscopy of SDSS J140737.17+442856.2, from the MaNGA survey, played an important role in the pre-selection of this target.  Unfortunately, WISE colours as red as SDSS J140737.17+442856.2 are rare - SDSS J140737.17+442856.2 is the only galaxy in the SDSS DR13 with $W1-W2 > 0.8$.  Nonetheless, there are good prospects for applying similar pre-selection strategies to a larger sample.  First of all, the $W1-W2 > 0.8$ can be somewhat relaxed -- simulations by Blecha et al. (in prep) suggest that a cut of $W1-W2>0.5$ actually identifies a higher fraction of AGN in the dual phase than a cut of $W1-W2 > 0.8$.  Indeed, half of the dual AGN identified by Satyapal et al. (2017) have $0.5 < W1-W2 < 0.8$.  Second, although galaxies with extremely red WISE colours may be rare in the MaNGA sample in general, luminous AGN (including a sub-set selected based on WISE colours) are included in the ancillary target list, greatly increasing the sample that can be studied with resolved spectroscopy.  Finally, we note that SDSS J140737.17+442856.2 could have been identified as a candidate dual AGN even in the absence of MaNGA IFU data - the SDSS Legacy (single fibre) and WISE surveys alone would have permitted the identification of two nuclei with complementary AGN signatures with a small separation.  The combination of WISE and optical AGN diagnostics may therefore make a powerful partnership for the identification of a larger sample of dual AGN that is complementary to techniques that identify the highly obscured phase.

\vspace{-0.7cm}
\section*{Acknowledgements}

This work would not have been possible without the tremendous ongoing effort of the SDSS collaboration, and the MaNGA team in particular.  We thank CXO for granting the DDT observations that enabled the prompt confirmation of this new dual AGN.  This research has made use of the NASA/IPAC Extragalactic Database (NED) which is operated by the Jet Propulsion Laboratory, California Institute of Technology, under contract with the National Aeronautics and Space Administration. This research made use of APLpy, an open-source plotting package for Python (Robitaille \& Bressert 2012).  This research made use of Astropy, a community-developed core Python package for Astronomy (Astropy collaboration et al. 2013).  


\vspace{-0.7cm}

\begin{thebibliography}{}
\small
\itemindent -0.48cm


\bibitem[Arnaud(1996)]{Arnaud96} Arnaud, K.~A.\ 1996, Astronomical Data Analysis Software and Systems V, 101, 17


\bibitem[Astropy Collaboration et al.(2013)]{AstropyCollaboration13} Astropy Collaboration, Robitaille, T.~P.,  et al.\ 2013, A\&A, 558, A33 

\bibitem[Ballo et al. (2004)]{bal04}
        Ballo, L., Braito, V., Della Ceca, R., Maraschi, L.,
	Tavecchio, F., Dadina, M.,  2004, ApJ, 600, 634

\bibitem[Bianchi et al. (2008)]{bia08}
        Bianchi, S., Chiaberge, M., Piconcelli, E., Guainazzi, M.,
	Matt, G., 2008, MNRAS, 386, 105

\bibitem[Brinchmann et al. 2004]{bri04} 
        Brinchmann, J., Charlot, S., White, S. D. M., Tremonti, C., 
	Kauffmann, G., Heckman, T., Brinkmann, J.,2004, MNRAS, 351, 1151  

\bibitem[Buchner et al. (2014)]{b14}
        Buchner, J., et al., 2014, A\&A, 564A, 125
        
\bibitem[Bundy et al. (2015)]{bun15}
        Bundy, K., et al., 2015, ApJ, 798, 7

\bibitem[Comerford et al. (2012)]{com12}
        Comerford, J. M., Gerke, B. F., Stern, D., Cooper, M. C., Weiner,
	B. J., Newman, J. A., Madsen, K., Barrows, R. S.,
	2012, ApJ, 753, 42

\bibitem[Comerford et al. (2015)]{com15}
         Comerford, J. M., Pooley, D., Barrows, R. S., Greene, J. E.,%
	 Zakamska, N. L., Madejski, G. M., Cooper, M. C.,
	  2015, ApJ, 806, 219

\bibitem[Ellison et al. (2013)]{post} 
         Ellison, S. L., Mendel, J. T.,  Patton, D. R., Scudder, J. M.,
	2013, MNRAS, 453, 3627

\bibitem[Ellison, Patton \& Hickox (2015)]{eph15}
        Ellison, S. L., Patton, D. R., Hickox, R. C., 2015, MNRAS, 451, L35

\bibitem[Ellison et al. (2008)]{sle08a}
         Ellison, S. L., Patton, D. R., Simard, L., McConnachie, A. W.,
	 2008 AJ, 135, 1877

\bibitem[Ellison et al. (2011)]{agn}
         Ellison, S. L., Patton, D. R.,  Mendel, J. T., Scudder, 
	 J. M., 2011, MNRAS, 418, 2043

\bibitem[Feroz \& Hobson (2008)]{fh08}
        Feroz, F., \& Hobson, M. P., 2008, MNRAS, 384, 449
     
\bibitem[Feroz et al. (2009)]{fer09}
        Feroz, F., Hobson, M. P., \& Bridges, M. 2009, MNRAS, 398, 1601

\bibitem[Feroz et al. (2013)]{fer13}
        Feroz, F., Hobson, M. P., Cameron, E., Pettitt, A. N., arXiv:1306.2144
       
\bibitem[Fruscione et al.(2006)]{Fruscione+06} Fruscione, A.,  et al.\ 2006, SPIE, 6270, 62701V 

\bibitem[Fu et al. (2012)]{fu12}
        Fu, H., Yan, L., Myers, A. D., Stockton, A., Djorgovski, S. G.,
	Aldering, G., Rich, J. A., 2012, ApJ, 745, 67

\bibitem[Jahnke \& Maccio (2011)]{jm11}
        Jahnke, K., \& Maccio, A. V., 2011, ApJ, 734, 92

\bibitem[Jarrett et al. (2011)]{j11}
        Jarrett, T. H., et al, 2011, ApJ, 735, 112        
        
\bibitem[Kauffmann et al. (2003)]{kau03} 
	Kauffmann, G., et al., 2003, MNRAS, 346, 1055

\bibitem[Kewley et al. (2001)]{kew01}  
         Kewley, L. J., Dopita, M. A., Sutherland, R. S., Heisler, C. A., 
	 Trevena, J.,  2001, ApJ, 556, 121


\bibitem[Khabiboulline et al. (2014)]{emil14}
        Khabiboulline, E. T., Steinhardt, C. L., Silverman, J. D.,
	Ellison, S. L., Mendel, J. T., Patton, D. R.,
	 2014, ApJ, 795, 62

\bibitem[Khan et al. (2016)]{khan16}
         Khan, F. M., Fiacconi, D., Mayer, L., Berczik, P., Just, A.,
	 2016, ApJ, 828, 73

\bibitem[Kocevski et al. (2015)]{koc15}
        Kocevski, D. D., et al.,  2015, ApJ, 814, 104

\bibitem[Komossa et al. (2003)]{kom03}
        Komossa, S., Burwitz, V., Hasinger, G., Predehl, P.,
	Kaastra, J. S., Ikebe, Y., 2003, ApJ, 582, L15

\bibitem[Lackner et al. (2014)]{lack14}
        Lackner, C. N., et al., 2014, ApJ, 148, 137

\bibitem[Law et al., (2016)]{law16}
        Law, D. R., et al., 2016, AJ, 152, 83

\bibitem[Lehmer et al. (2010)]{leh10}
        Lehmer, B. D., et al., 2010, ApJ, 724, 559
        
\bibitem[Liu et al. 2013]{liu13} 
        Liu, X., Civano, F., Shen, Y., Green, P., Greene, J. E.,
	Strauss, M. A.,  2013, ApJ, 762, 110

\bibitem[Maiolino et al.(2001)]{Maiolino+01}
        Maiolino, R.,  et al.\ 2001, A\&A, 365, 28 

        
\bibitem[Marchesi et al.(2016)]{Marchesi+16} Marchesi, S.,  et al.\ 2016, ApJ, 830, 100 

\bibitem[Mendel et al. (2014)]{mass}
        Mendel, J. T., Palmer, M. J. D., Simard, L., Ellison, S. L.,  
	Patton, D. R., 2014, ApJS, 210, 3


\bibitem[Park et al.(2006)]{Park+06} Park, T.,  et al.\ 2006, ApJ, 652, 610 

\bibitem[Patton et al. (2016)]{dave16}
         Patton, D. R., Qamar, F. D., Ellison, S. L.,  Bluck, A. F. L.,
	 Simard, L., Mendel, J. T., Moreno, J., Torrey, P.,
	 2016, MNRAS, 461, 2589

\bibitem[Ricci et al. (2017)]{ric17}
        Ricci, C., et al., 2017, MNRAS, in press

\bibitem[Ricci et al. (2015)]{bat}
        Ricci, C., Ueda, Y., Koss, M. J., Trakhtenbrot, B., Bauer, F. E.,
	Gandhi, P.,  2015, ApJ, 815, L13

\bibitem[Richards et al. (2009)]{r09}
         Richards, J. W., Freeman, P. E., Lee, A. B., Schafer, C. M.,
         2009, MNRAS, 399, 1044

\bibitem[Robitaille \& Bressert(2012)]{RobitailleBressert12} Robitaille, T., \& Bressert, E.\ 2012, ASCL, ascl:1208.017

\bibitem[Sanchez-Blazquez et al. (2006)]{sb06}
        Sanchez-Blazquez, P., et al., 2006, MNRAS, 371, 703
  
\bibitem[Satyapal et al. (2014)]{sat14} 
        Satyapal, S., Ellison, S. L., McAlpine, W., Hickox, R. C., 
	Patton, D. R., Mendel, J. T., 2014, MNRAS, 441, 1297

\bibitem[Satyapal et al. (2017)]{sat17} 
        Satyapal, S.,  et al., 2017, MNRAS, submitted

\bibitem[Sesana et al. (2004)]{ses04}
        Sesana, A., Haardt, F., Madau, P., Volonteri, M.,
	 2004, ApJ, 611, 623

\bibitem[Scudder et al. (2012b)]{[pairs}
         Scudder, J. M., Ellison, S. L., Torrey, P., Patton, D. R.,
	 Mendel, J. T., 2012, MNRAS, 426, 549

\bibitem[Simard et al. (2011)]{sim11}
        Simard, L., Mendel, J. T., Patton, D. R., Ellison S. L., 
	McConnachie, A. W., 2011, ApJS, 196, 11

\bibitem[Stern et al. (2012)]{ste12}
         Stern, D., et al., 2012, ApJ, 753, 30

\bibitem[Teng et al (2012)]{te12}
        Teng, S. H., et al.,  2012, ApJ, 753, 165

\bibitem[Willingale et al.(2013)]{Willingale+13} Willingale, R., Starling, R.~L.~C., Beardmore, A.~P., Tanvir, N.~R., \& O'Brien, P.~T.\ 2013, MNRAS, 431, 394 

\bibitem[Wright et al. (2010)]{wise}
        Wright, E. L., et al., 2010, AJ, 140, 1868

\end{thebibliography}
\end{document}